\documentclass{article}
\usepackage{spconf,amsmath,amssymb,graphicx,booktabs,multirow,microtype,url}
\usepackage{tikz}
\usetikzlibrary{arrows.meta,calc,positioning}
\usepackage[T1]{fontenc}

\usepackage{etoolbox}
\patchcmd{\thebibliography}
  {\settowidth}
  {\setlength{\itemsep}{0pt}\setlength{\parskip}{0pt}\settowidth}
  {}{}

\title{PIXEL-KWS: Rendered-Glyph Representations for\\
       Orthographically Robust Open-Vocabulary Keyword Spotting}

 \name{Adarsh Arigala$^{1}$, Ramesh Gundluru$^{2,3}$, Srinivasan Umesh$^{1}$, Ludovick Lepauloux$^{3}$, Hakim Hacid$^{3}$ }

\address{
$^{1}$ SPRING Lab, Indian Institute of Technology Madras, India \\
$^{2}$ Speech Information Processing Lab, Indian Institute of Technology Hyderabad, India \\
$^{3}$ Technology Innovation Institute, UAE 
}
 
\begin{document}
\maketitle
 
\begin{abstract}

Open-vocabulary keyword spotting (KWS) allows users to enroll arbitrary keywords at inference time. However, character and phoneme-based text encoders rely on fixed symbol vocabularies learned during training, limiting their ability to handle unseen or visually similar symbols such as homoglyphs, l33tspeak substitutions, and accented characters. We propose PIXEL-KWS, an open-vocabulary KWS framework that represents characters through rendered glyphs rather than learned embedding lookups. We render each character as a grayscale image and encode it with a lightweight CNN, enabling the text encoder to generalize from visual form rather than a fixed vocabulary. On LibriPhrase, PIXEL-KWS and its matched character-based counterpart achieve $9.57\%$ and $9.55\%$ EER on LibriPhrase-hard,  $0.41\%$ and $0.43\%$ EER on LibriPhrase-easy, outperforming all published baselines without pre-training or synthetic data. PIXEL-KWS provides substantially greater robustness to both homoglyph and l33tspeak symbol-level perturbations. These results highlight glyph-based representations as a simple and effective approach for robust open-vocabulary KWS in practical settings.

\end{abstract}


\section{Introduction}
\label{sec:intro}

Open-vocabulary keyword spotting (KWS) matches speech with a keyword supplied by the user as text, allowing new keywords to be enrolled without additional speech examples \cite{baseline_cmcd,baseline_phonmatchnet,baseline_ced,plcl}. The text encoder operates directly on user-provided input, making its tokenizer is important design. Most existing approaches use either character-based or phoneme-based tokenizers. Character-based or grapheme-to-phoneme (G2P) tokenizers  typically represent each symbol using a learnable embedding lookup table, resulting in a closed vocabulary. This limitation is particularly relevant to practical user input, where keywords may contain homoglyphs, l33tspeak substitutions, accented characters, or other unseen symbols as shown in Fig:\ref{fig:render}. Such symbols can naturally arise through copy-and-paste or other forms of user input, exposing a fundamental limitation of fixed-vocabulary.


To remove limitations on fixed vocabulary, NLP research has explored representing text through its visual form. Salesky et al.~\cite{salesky2021visual} render source text as images for machine translation and show improved robustness to orthographic noise. PIXEL~\cite{rust2023pixel} further pre-trains a language model directly on rendered text, enabling transfer across scripts without relying on a fixed textual vocabulary. Similar visual text representations have also been explored in speech synthesis as alternatives to conventional character representations \cite{vTTS,VisualOnomaToWave,Adarsh2026pixeltts}. While Pixel-TTS \cite{Adarsh2026pixeltts} uses image-based text rendering for speech synthesis, its robustness is evaluated using speech quality and intelligibility measures under textual noise. In contrast, open-vocabulary KWS requires keyword representations that remain matchable to audio under enrollment-time symbol corruption. These studies suggest that visual representations can improve robustness to textual variation and generalize beyond symbols seen during training~\cite{Pixel-level-Fallback}.

 To the best of our knowledge, no open-vocabulary KWS system directly represents user-provided keywords as rendered text images. We introduce \textbf{PIXEL-KWS}, an open-vocabulary KWS framework that renders each character of a keyword as a $16\times16$ grayscale image and encodes the resulting sequence with a lightweight CNN, replacing the conventional character-embedding lookup table. Our contributions are:

\begin{itemize}
\itemsep1pt
\item We introduce image-rendered glyph representations for open-vocabulary KWS, while retaining $O(1)$ keyword enrollment.
\item Without any synthetic data or pre-training, both our systems outperform all published baselines ($9.55\%$ and $9.57\%$ EER on LP\textsubscript{H}), so the robustness comparison is made against a strong rather than a weak one.
\item We evaluate robustness to orthographic noise by introducing homoglyph and l33tspeak corruptions at varying corruption probabilities.
\item We show that the gains are particularly evident for visually similar substitutions: under full homoglyph corruption, PIXEL-KWS degrades by only $11.2\%$ EER, compared with $31.3\%$ for the character baseline.
\item We systematically evaluate character image resolutions of $8\times8$, $16\times16$, and $24\times24$, showing that resolution affects KWS performance.
\end{itemize}

\section{PIXEL-KWS Methodology}
\label{sec:method}

\begin{figure}[t]
\centering
\includegraphics[width=\linewidth]{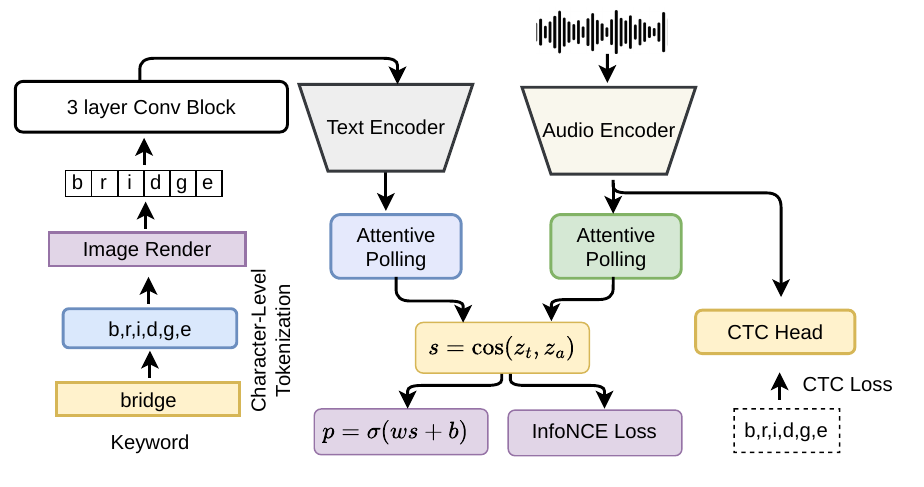}
\caption{Overview of the proposed PIXEL-KWS framework.}
\label{fig:arch}
 
\end{figure}

\subsection{Character-to-Image representation}
\label{ssec:render}

Let $y=(y_1,\dots,y_L)$ denote the sequence of characters in the enrolment keyword. Each character is rendered a grayscale character image called glyph $G_c \in [0,1]^{16\times16}$ using the bold monospaced DejaVu Sans Mono font. Each character $c$ is cached, and indexed by character ID to avoid repeated rendering. The word-break character (space) is represented by a white-space image and distinguished from padding using the sequence mask. This preserves word boundaries in multi-word keywords (phrases), preventing distinct keywords such as \emph{``a bridge''} and \emph{``abridge''} from becoming visually indistinguishable. 

These glyph representations are processed by three convolutional layers with channel dimensions $1\rightarrow8\rightarrow8\rightarrow16$, with stride 2 in the first two layer. Each convolutional layer is followed by GroupNorm. The resulting $16\times4\times4$ feature map is flattened and linearly projected to a $64$-dimensional character representation. The representations are then processed by a bidirectional GRU to capture contextual dependencies:
\vspace{-0.1cm} 
\begin{equation}
e_l = \mathrm{CNN}_{\theta}(G_{y_l}) \in \mathbb{R}^{d},
\quad
z_{1:L} = \mathrm{BiGRU}(e_{1:L}),
\end{equation}
where $e_l$ denotes the representation of the $l$-th character and $z_l$ its contextualized representation. We then aggregate these representations using masked attentive pooling:
\begin{equation}
z = \sum_{l=1}^{L} \alpha_l z_l,
\end{equation}
where $\alpha_l$ is the attention weight assigned to the $l$-th character.The pooled representation is  linearly projected into the shared $D$-dimensional latent space to obtain the keyword embedding $z_y\in\mathbb{R}^{D}$.

\subsection{Audio Keyword representation}
The audio encoder consists of a $4\times$ convolutional subsampler followed by a $4$-layer Conformer~\cite{gulati2020conformer}, which takes log-Mel spectrograms as input and produces frame-level representations. These representations are aggregated using masked attentive pooling and linearly projected into the shared $D$-dimensional latent space to obtain the embedding of audio keyword $z_a\in\mathbb{R}^{D}$.

\subsection{Training Objective}
Given $N$ paired audio-text keyword embeddings, we compute their pairwise cosine similarities to form an $N\times N$ similarity matrix, where matched pairs lie on the diagonal and all other pairs act as in-batch negatives. We train the model with a bidirectional InfoNCE objective at temperature $\tau$, considering both audio-to-text and text-to-audio matching:

\begin{equation}
\mathcal{L}_{\mathrm{InfoNCE}}
=
\frac{1}{2}
\left(
\mathcal{L}_{a\rightarrow t}
+
\mathcal{L}_{t\rightarrow a}
\right).
\end{equation}

 We add a CTC head to the Conformer output to jointly predict the phoneme sequence corresponding to the audio keyword. We further calibrate the cosine similarity using a learnable affine transformation followed by a sigmoid. 

\begin{equation}
p = \sigma\!\left(w\,s_{at} + b\right),
\end{equation}
where $w$ is reparameterized using a softplus function to ensure a positive scale. The overall objective combines the contrastive, CTC, and binary cross-entropy losses:
 
\begin{equation}
\mathcal{L}
=
\mathcal{L}_{\mathrm{InfoNCE}}
+
\lambda_{\mathrm{ctc}}\mathcal{L}_{\mathrm{CTC}}
+
\lambda_{\mathrm{bce}}\mathcal{L}_{\mathrm{BCE}},
\label{eq:head}
\end{equation}
where $\mathcal{L}_{\mathrm{BCE}}$ is computed using the pair label, with $y=1$ for matched audio--text pairs and $y=0$ for negative pairs.

\section{Experimental Setup}
\label{sec:setup}

\textbf{Data.} We train our models using LibriPhrase dataset\cite{baseline_cmcd}, derived from the LibriSpeech \emph{train-clean-100} and \emph{train-clean-360} subsets \cite{panayotov2015librispeech}. We evaluate the proposed models on official LibriPhrase test set constructed from \emph{train-other-500}, under easy (LP\textsubscript{E}) and hard (LP\textsubscript{H}) conditions. We hold out $10\%$ of the training non-overlapping keyword trials as validation data for model selection. We additionally mine hard negatives for each keyword and include them during training. Following the reference evaluation protocols \cite{baseline_phonmatchnet,ai24_interspeech}, each test pair is evaluated in both directions, with the anchor audio matched against comparison text and the anchor text matched against comparison audio. We report both Equal Error Rate (EER) and Area Under the Receiver Operating Characteristic (AUROC) for each evaluation condition. The EER is computed at the operating point where the false acceptance rate and false rejection rate are equal, while AUROC summarizes the system's discrimination performance across all possible decision thresholds.

\textbf{Model and training.} The audio encoder uses $d_{\mathrm{model}}=64$ with 4 Conformer layers, 4 attention heads, and an FFN dimension of 128. The text encoder uses a 2-layer BiGRU with hidden size 128, and both modalities are projected into a 64-dimensional matching space. The character baseline
(Char-KWS) has 0.847M parameters, while PIXEL-KWS has
0.863M parameters. We intentionally designed Glyph CNN module to be lightweight, so that performance differences cannot be attributed simply to increased model capacity. All the models are trained for 30 epochs using AdamW with a learning rate of $10^{-3}$ and 2000 warmup steps, with a batch size of 128 per GPU across 4 GPUs using \textit{bf16} precision.

\vspace{-0.2cm} 
\section{Results and analysis}
\label{sec:results}

\begin{table}[t]

\caption{Comparison of open-vocabulary KWS methods on LibriPhrase. Char-KWS and PIXEL-KWS are both ours and differ only in the first stage of the text encoder; T/S/M denote $d_{\mathrm{model}}$ $64$/$128$/$256$. IS: Interspeech; ICP: ICASSP ; $*$ ASR-based pre-training, $\dagger$
phrase-based, $\ddagger$ synthetic data.}
\label{tab:main_table_res}
\centering
\resizebox{\columnwidth}{!}{%
\begin{tabular}{@{}llc c c c c@{}}
\hline
\multirow{2}{*}{\textbf{Method}} &
\multirow{2}{*}{\textbf{Venue}} &
\multirow{2}{*}{\textbf{\#Params}} &
\multicolumn{2}{c}{\textbf{AUROC (\%)} $\uparrow$} &
\multicolumn{2}{c}{\textbf{EER (\%)} $\downarrow$} \\
\cline{4-5}\cline{6-7}
& & &
\textbf{LP}$_\textbf{H}$ & \textbf{LP}$_\textbf{E}$ &
\textbf{LP}$_\textbf{H}$ & \textbf{LP}$_\textbf{E}$ \\
\hline

\multicolumn{7}{@{}l}{\textit{Posterior-based}} \\
DONUT~\cite{lugosch2018donut} & ICP 21 & --- & 62.55 & 78.74 & 41.95 & 28.74 \\
CTCAT~\cite{jin24d_interspeech} & IS 24 & 0.2M & 77.10 & 98.32 & 29.63 & 6.06 \\
W-CTC~\cite{kim25d_interspeech} & IS 25 & 3.6M & 95.93 & 99.95 & 10.21 & 0.91 \\

\hline
\multicolumn{7}{@{}l}{\textit{Cross-attention-based}} \\
CMCD~\cite{baseline_cmcd} & IS 22 & 0.7M & 73.58 & 96.70 & 32.90 & 8.42 \\
PhonMatchNet~\cite{baseline_phonmatchnet} & IS 23$*$ & 0.7M & 88.52 & 99.29 & 18.82 & 2.80 \\
CED~\cite{baseline_ced} & ICP 24$\dagger$ & 3.6M & 92.70 & 99.84 & 14.40 & 1.70 \\
MM-KWS@T~\cite{ai24_interspeech} & IS 24$\ddagger$ & 3.9M & 95.36 & 99.94 & 10.41 & 0.82 \\
SLiCK~\cite{slick} & ICP 25$\dagger$ & 0.6M & 94.90 & 99.82 & 11.10 & 1.78 \\
PLCL@T~\cite{plcl} & ICP 25$\dagger$ & 40M & 95.56 & 99.95 & 9.96 & 1.21 \\

\hline
\multicolumn{7}{@{}l}{\textit{Embedding-based}} \\
Triplet~\cite{baseline_triplet} & IS 19 & --- & 54.88 & 63.53 & 44.36 & 32.75 \\
CLAD~\cite{baseline_clad} & ICP 24 & 3.6M & 76.15 & 97.03 & 30.30 & 8.65 \\
SynapSpot-TA~\cite{synapspot} & ICP 26 & 0.9M & 79.15 & 97.34 & 27.29 & 5.77 \\
ADML~\cite{baseline_adml} & IS 25$*$ & 1.8M & 88.71 & 99.86 & 20.09 & 1.33 \\
MATE~\cite{mate} & ICP 26 & --- & 88.70 & 99.86 & 20.06 & 1.38 \\

Char-KWS-T (this work) & --- & 0.85M & 94.41 & 99.97 & 12.40 & 0.63 \\
PIXEL-KWS-T (this work) & --- & 0.86M & 94.65 & 99.97 & 12.10 & 0.60 \\
Char-KWS-S (this work) & --- & 2.9M & 95.96 & 99.98 & 10.11 & 0.39 \\
PIXEL-KWS-S (this work) & --- & 3.0M & 95.91 & 99.98 & 10.22 & 0.41 \\
\textbf{Char-KWS-M} (this work) & --- & 11.4M & \textbf{96.27} & \textbf{99.98} & \textbf{9.55} & \textbf{0.43} \\
\textbf{PIXEL-KWS-M} (this work) & --- & 11.4M & \textbf{96.25} & \textbf{99.98} & \textbf{9.57} & \textbf{0.41} \\

\hline
\end{tabular}%
}
\vspace{-0.4cm}
\end{table}

\subsection{Performance Analysis}
\label{ssec:main}

Table~\ref{tab:main_table_res} shows PIXEL-KWS performance on LibriPhrase at three model sizes. The tiny variant (T, $0.86$M parameters) achieves 12.10\% EER and 94.65\% AUROC on LP\textsubscript{H}, and 0.60\% EER and 99.97\% AUROC on LP\textsubscript{E}, the best result among the embedding-based systems despite its substantially smaller size and the absence of pre-training. Increasing the encoder dimension $d_{\mathrm{model}}$ to 128 (S) and 256 (M), improves LP\textsubscript{H} EER to 10.22\% and 9.57\%. At the medium size both of our systems outperform every posterior-based, cross-attention-based and embedding-based baselines. Char-KWS-M reaches 9.55\% EER and 96.27\% AUROC on LP\textsubscript{H} and PIXEL-KWS-M 9.57\% and 96.25\%, against PLCL@T, which uses $680k$h of pre-training. Whereas our systems do not use any pre-training and synthetic data. Char-KWS is not a prior system but our own matched baseline, identical to PIXEL-KWS except for the first stage of the text encoder. This controlled comparison isolates the effect of the glyph-based representation while keeping the architecture and training fixed. On clean trials the two front-ends are equivalent at every size (12.40 versus 12.10\%, 10.11 versus 10.22\%, and 9.55 versus 9.57\%).

A further advantage of our embedding based model is the enrolment cost. Because the text side is reduced to a single embedding, a keyword is encoded once and scored by an inner product, which is $O(1)$ per enrolled keyword and need not be recomputed. Cross-attention systems fuse the audio and text representations and therefore repeat the computation for every enrolled keyword, while posterior-based systems run a per-keyword decoding search. PIXEL-KWS provides substantially improved robustness to orthographic variations and character-level perturbations, as discussed in the following subsection \ref{ssec:robust}.

\vspace{-0.4cm}
\subsection{Orthographic robustness}
\label{ssec:robust}

\begin{table}[t]
\centering
\footnotesize
\caption{EER (\%) under enrolment-string corruption at probability $p$; audio remains unchanged. $\Delta$ denotes degradation from $p=0$ to $p=1$. Best results are shown in bold.}

\label{tab:robust}
\setlength{\tabcolsep}{3.0pt}
\begin{tabular}{@{}llccc c ccc@{}}
\hline
\multirow{2}{*}{\textbf{Cond.}} & \multirow{2}{*}{\textbf{System}} &
\multicolumn{3}{c}{\textbf{homoglyph}} & & \multicolumn{3}{c}{\textbf{l33tspeak}} \\
\cmidrule(lr){3-5}\cmidrule(lr){7-9}
 & & $p=0$ & $p=1$ & $\Delta$ & & $p=0$ & $p=1$ & $\Delta$ \\
\hline
\multirow{2}{*}{LP$_\mathrm{E}$} & Char-KWS-M & 0.43 & 25.68 & +25.24 &  & 0.43 & 13.13 & +12.70 \\
& PIXEL-KWS-M & \textbf{0.41} & \textbf{2.65} & \textbf{+2.24} &  & \textbf{0.41} & \textbf{9.31} & \textbf{+8.90} \\
\hline
\multirow{2}{*}{LP$_\mathrm{H}$} & Char-KWS-M & \textbf{9.55} & 40.85 & +31.31 &  & \textbf{9.55} & 36.01 & +26.46 \\
& PIXEL-KWS-M & 9.57 & \textbf{20.81} & \textbf{+11.23} &  & 9.57 & \textbf{31.65} & \textbf{+22.07} \\
\hline
\end{tabular}
\vspace{-0.4cm}
\end{table}

We evaluate robustness of PIXEL-KWS using character-level perturbations. This setting is particularly relevant because visual similarity between original and corrupted characters can preserve keyword representations under orthographic noise. In contrast, phoneme or BPE-based representations convert corrupted characters into different phoneme or subword sequences and do not directly preserve this visual relationship. We consider two perturbations following \cite{salesky2021visual,Adarsh2026pixeltts}: Homoglyphs and L33tspeak. Homoglyphs are characters from different scripts that look nearly identical but have different Unicode values, whereas L33tspeak replaces standard letters with visually similar numbers.

\begin{figure}[t]
  \centering
  \includegraphics[width=\linewidth,height=3cm ]{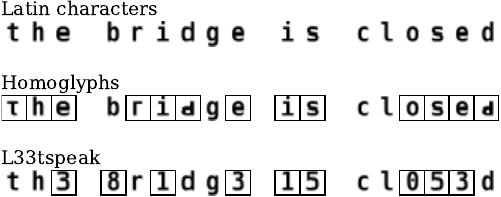}
  \caption{Visualization of the enrolment string under character noise, homoglyph, and l33tspeak corruption.}
  \label{fig:render}
 \vspace{-0.3cm}
\end{figure}

\begin{figure}[!h]
  \centering
  \includegraphics[width=\linewidth ]{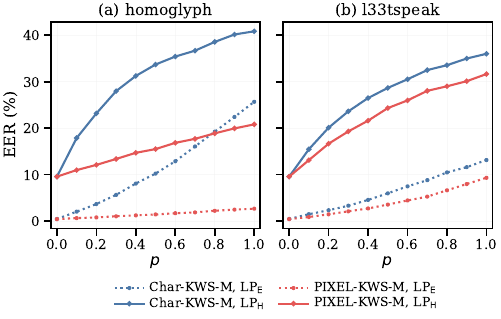}
\caption{EER against enrolment-string corruption probability $p$. Homoglyph and l33tspeak substitutions contain characters outside the training corpus.}
\label{fig:robust}
\end{figure}

Each eligible character of the enrolment keyword is substituted independently with probability $p$, and the audio is left unchanged, so any change in EER is attributable to the text encoder alone. Ten homoglyph and eight l33tspeak substitutions are defined, so at $p=1$ the two maps affect 67\% and 42\% of all characters. Fig.~\ref{fig:render} shows both corruptions applied to \emph{``the bridge is closed''}, with boxes marking the substituted positions: 13 positions in the homoglyph row and 9 in the l33tspeak row. The boxed homoglyphs are visually almost indistinguishable from the letters they replace, whereas the boxed l33tspeak characters are clearly digits. Fig.~\ref{fig:robust} present EERs across the full range of $p$, and Table~\ref{tab:robust} reports the degradation from $p=0$ to $p=1$. Under homoglyph corruption Char-KWS-M degrades by $31.316\%$  on LP\textsubscript{H} and $25.246\%$ points on LP\textsubscript{E}. But PIXEL-KWS-M degrades only $11.236\%$ and $2.246\%$ on LP\textsubscript{H} and LP\textsubscript{E} respectively . Under l33tspeak the corresponding degradations are $26.46\%$ and $12.70\%$ EERs for Char-KWS-M, and $22.07\%$ and $8.9\%$ EERs for PIXEL-KWS-M. Across the sweep, PIXEL-KWS-M obtains a lower EER than Char-KWS-M at every non-zero corruption probability, for both perturbation types and on both LP\textsubscript{E} and LP\textsubscript{H}.

\vspace{-0.5cm}

\subsection{Ablation: Character Image Resolutions}
\label{ssec:control}
Table~\ref{tab:res} shows that PIXEL-KWS is relatively insensitive to the character image resolution. The $16\times16$ resolution achieves the best performance on LP\textsubscript{H}, while $8\times8$ and $24\times24$ yield comparable results. Lower resolutions may discard fine-grained character-shape information, whereas higher resolutions introduce additional pixels that primarily represent white space around the glyph, increasing the computational and parameter overhead without providing a consistent performance gain. We therefore use $16\times16$ character images in all experiments, as it provides a good balance between retaining character details and computational efficiency.
\vspace{-0.3cm} 
\begin{table}[!h]
\centering
\caption{Performance with different character image resolutions }
\label{tab:res}
\setlength{\tabcolsep}{8pt}
\begin{tabular}{@{}llcc@{}}
\hline
\textbf{Image} & \textbf{\#Params} & \textbf{LP}$_\textbf{E}$ & \textbf{LP}$_\textbf{H}$ \\
\hline
$8\!\times\!8$ & 0.85M & 0.58 & 12.28 \\
$16\!\times\!16$ & 0.86M & 0.60 & 12.10 \\
$24\!\times\!24$ & 0.89M & 0.58 & 12.25 \\
\hline
\end{tabular}
\vspace{-0.2cm}
\end{table}

\subsection{Visual Embeddings for Unseen Symbols}
\label{ssec:tsne}
\begin{figure}[!h]
  \centering
  \includegraphics[width=0.6\linewidth,height=4cm]{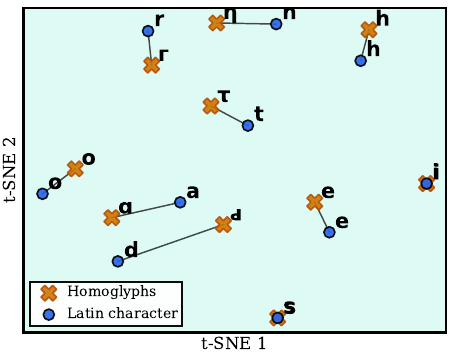}
  \caption{t-SNE visualisations of the character embeddings trained from PIXEL-KWS.}
  \label{fig:tsne}
\end{figure}

Fig.~\ref{fig:tsne} visualizes the t-SNE embeddings for characters. We extract the character embeddings of the original Latin characters and their homoglyph variants from the PIXEL-KWS model and project them into a 2-D t-SNE space. The resulting embeddings show that visually similar characters are mapped close to each other. Although the Unicode characters are different, PIXEL-KWS places the homoglyphs close to their corresponding Latin characters, with a mean cosine similarity of $+0.785$, compared with $-0.002$ between different Latin characters. The close correspondence arises because the representations are derived from the rendered glyphs, allowing visually similar characters to produce similar embeddings despite having different Unicode identities. In contrast, the character baseline has no embedding for these unseen symbols.

\vspace{-0.5cm} 

\section{Conclusion}
\label{sec:concl}

We presented \textbf{PIXEL-KWS}, an open-vocabulary keyword spotting framework that represents keyword characters through rendered glyphs rather than a fixed embedding vocabulary. This glyph representation allows the text encoder to process unseen symbols while preserving visual similarities between orthographically related characters. On LibriPhrase, PIXEL-KWS achieves performance comparable to the character-based baseline on clean keywords while providing substantially improved robustness to character-level perturbations. The glyph representation is script-independent by construction; extending PIXEL-KWS to multilingual settings is natural future work. Overall, the results demonstrate that visual character representations are a simple and effective approach for improving the robustness of open-vocabulary KWS to unseen and visually confusable symbols.


\bibliographystyle{IEEEbib}
\bibliography{refs}

\end{document}